# ARTEMIS observations of electrostatic shocks inside the lunar wake

Terry Z. Liu[1*], Xin An[1*], Vassilis Angelopoulos[1], and Andrew R. Poppe[2]

[1]Department of Earth, Planetary, and Space Sciences, University of California, Los Angeles, USA. [2]Space Sciences Laboratory, University of California, Berkeley, Berkeley, CA, USA

**Abstract**

When the solar wind encounters the Moon, a plasma void forms downstream of it, known as the lunar wake. In regions where the magnetic field is quasi-parallel to the plasma-vacuum boundary normal, plasma refills the wake primarily along magnetic field lines. As faster electrons outpace slower ions, an ambipolar electric field is generated, accelerating ions and decelerating electrons. Recent particle-in-cell simulations have shown that when accelerated supersonic ion beams from opposite sides of the wake meet near the wake center, electrostatic shocks may form, decelerating ions and heating electrons into flat-top velocity distributions. Using data from the Acceleration, Reconnection, Turbulence and Electrodynamics of the Moon's Interaction with the Sun (ARTEMIS) spacecraft, we present the first observational evidence of the predicted electrostatic shocks. Near the wake center of one event, we observed an electrostatic solitary structure with an amplitude of ~2 mV/m and a spatial scale of ~50 local Debye lengths. This structure generated a potential increase of ~50 V from upstream to downstream, heating incoming electrons by ~50 eV in the parallel direction while decelerating ions by ~60 km/s leading to a density enhancement. At a second event representing a more evolved stage, we observed more dissipated structures dominated by strong electrostatic waves, with persistent potential increases driving continued field-aligned electron heating and ion deceleration. These observations confirm simulation predictions of electrostatic shock formation and the associated particle dynamics within the lunar wake, with potential applications to understanding plasma interactions around other airless celestial bodies.

*Correspondence to: Terry Z. Liu terryliuzixu@ucla.edu
Xin An phyax@ucla.edu

## 1. Introduction

When a supersonic plasma flow encounters an unmagnetized, airless body, the upstream-facing surface absorbs most of the plasma, creating a plasma void on the downstream side. The lunar wake serves as a typical example of this phenomenon, offering accessible in-situ spacecraft observations that provide a natural laboratory for understanding plasma expansion into vacuum (see review by Halekas et al., 2015). The process of solar wind plasma refilling the lunar wake depends strongly on the orientation of the interplanetary magnetic field (IMF). When the IMF is quasi-perpendicular to the wake boundary normal, fluid dynamics governs the refilling process. As plasma expands inward across magnetic field lines at the wake boundary, the magnetic field in the wake becomes compressed, associated with diamagnetic currents at the boundary (e.g., Colburn et al., 1967; Halekas et al., 2005). At the same time, a rarefaction wave propagates outward from the boundary at the magnetosonic speed (e.g., Zhang et al., 2012, 2014).

When the IMF is quasi-parallel to the wake boundary normal, kinetic processes along magnetic field lines become dominant. As solar wind electrons move along field lines into the wake much faster than ions, ambipolar electrostatic fields develop to accelerate ions into the wake and maintain quasi-neutrality. Theoretical studies have derived self-similar solutions for one-dimensional (1-D) plasma expansion by combining ion fluid equations with isothermal electrons following the Boltzmann relation under the quasi-neutral condition (Denavit 1979; Samir et al. 1983; Halekas et al., 2014). In these solutions, the field-aligned ion speed follows $V_i = s/t + C_s$, where $s$ is the distance along field lines, $t$ is the expansion time (determined from solar wind convection time along the wake boundary surface in observations), and $C_s$ is the ion acoustic speed.

When accelerated ion beams from opposite sides of the wake encounter each other, electrostatic ion-ion instabilities excite electrostatic waves (e.g., Stringer, 1964; Farrell et al., 1997). These waves can steepen into solitons and electrostatic shocks that decelerate incoming ions while heating incoming electrons. During the initial overlap of the two counter-streaming ion beams, the electron thermal pressure gradient at the beam front drives an electrostatic field that decelerates incoming ions (e.g., Dieckmann et al., 2014). When the incoming ion beams are faster than the local ion acoustic speed, electrostatic shocks develop (e.g., Forslund and Shonk, 1970; Dieckmann et al., 2014).

Because the theoretical field-aligned ion speed $V_i$ in the wake always exceeds the local ion acoustic speed, electrostatic shocks are expected to form. Recent 1-D particle-in-cell (PIC) simulations by An et al. (2025) identified such electrostatic shocks at the wake center. These shocks operate through an electrostatic field pointing from downstream to upstream, decelerating incoming ions and creating a density enhancement in the downstream region while heating electrons in the field-aligned direction (e.g., Forslund and Shonk, 1970). Electrostatic shocks thermalize ions indirectly through other instabilities, such as the electrostatic ion-ion instability between incoming ions and shock-reflected ions (e.g., Davidson et al., 1970; Forslund and Freidberg, 1971). The characteristic spatial scale of electrostatic shocks spans a few tens of local Debye lengths (An et al., 2025) or approximately one electron skin depth (e.g., Kato and Takabe, 2010; Dieckmann et al., 2014).

Although electrostatic shocks have been identified in both simulations and laboratory experiments (e.g., Karimabadi et al., 1991; Sakawa et al., 2016), direct observations in space remain very rare, occurring mainly in the auroral zones (e.g., Mozer, 1981). This scarcity may be due to their characteristically small spatial scales, which require unrealistically high-resolution plasma measurements to distinguish electrostatic shocks from other electrostatic structures such as double layers. The lunar wake environment, however, presents a unique opportunity for such observations. The extremely low plasma density within the wake produces very large electron skin depths (tens of km) and Debye lengths (~1 km), substantially increasing the likelihood that spacecraft can resolve the associated plasma dynamics. Using the Acceleration, Reconnection, Turbulence and Electrodynamics of the Moon's Interaction with the Sun (ARTEMIS) spacecraft, we report the observations of electrostatic shocks and their associated particle dynamics inside the lunar wake. These observations are consistent with predictions from recent simulations (An et al., 2025).

## 2. Data

Two ARTEMIS spacecraft, P1 and P2, were part of the Time History of Events and Macroscale Interactions during Substorms (THEMIS, Angelopoulos, 2008) mission before 2010 and were subsequently placed into lunar orbits in the equatorial plane. For this study, we analyze plasma data from the electrostatic analyzer (ESA) at ~4s resolution (McFadden et al., 2008) and magnetic field data from the fluxgate magnetometer (FGM) at ~0.0625s resolution (Auster et al., 2008). Electric field data from the electric field instrument (EFI; Bonnell et al., 2009), also at ~0.0625s resolution, are analyzed in the spacecraft spin plane, which is approximately aligned with the equatorial plane, as measurements along the spin axis lack reliable calibration (contaminated by the spacecraft potential). The spacecraft potential from EFI is used to infer electron density at ~0.125s resolution by fitting a Boltzmann relation.

We select events where one spacecraft traverses the lunar wake and the other monitors the background solar wind conditions. To observe interactions between counter-streaming ion beams along magnetic field lines in the equatorial plane, we require periods of dominant and stable IMF

$B_y$ component in Selenocentric Solar Ecliptic (SSE) coordinates. We present two events in the main text and include two additional events in the supplementary materials (Figures S1 and S2).

To compare our observations with the ion refilling model $V_i$ (Denavit 1979; Samir et al. 1983; Halekas et al., 2014), we trace the locally measured magnetic field directions forward and backward from the spacecraft position to the lunar wake boundaries, assuming a cylindrical wake geometry with radius equal to the lunar radius. Based on the intersection points where magnetic field lines cross the wake boundaries, we calculate two key parameters: the distance $s$ along field lines to the spacecraft, and the distance $L$ from the Moon's center to the boundary intersection along the SSE-X axis. Using the background solar wind speed $V_{sw}$, we obtain the expansion time $t = L/V_{sw}$. We calculate the ion acoustic speed $C_s = \sqrt{T_e/m_i}$ using the locally measured electron temperature $T_e$ and the proton mass $m_i$. We then obtain $V_i$ from both sides of the wake in the solar wind rest frame and $\vec{V}_i + \vec{V}_{sw}$ in the spacecraft frame.

## 3. Results

### 3.1. Event 1

In this event, P2, located in the background solar wind, observed an IMF dominated by a negative SSE-Y component (Figure 1a; blue arrow in Figure 1h). Simultaneously, P1 traversed the lunar wake from dawnside to duskside (red arrow in Figure 1h). We observed two distinct ion beams originating from opposite sides of the lunar wake – they are well described by the ion refilling model (black and red lines in Figure 1e). The dawnside ion beam exhibited higher energy than the duskside beam, due to the background solar wind velocity component along the tilted IMF.

Shortly after 04:20 UT, when the two ion beams almost collided (due to low counts, it is difficult to determine whether they had overlapped), a localized density enhancement appeared near the wake center (black arrow in Figure 1c; red star in Figure 1h). This feature was accompanied by an increase in ion energy flux at the energies corresponding to the duskside beam (Figure 1e), enhanced electron energy flux (Figure 1f), and a peak in the electron parallel temperature (Figure 1g). Taken together, these observations suggest the formation of a plasma structure resulting from ion beam–beam interaction. We next examine this structure in greater detail in Figure 2.

Figure 2b shows that the electron density (measured from ESA, $N_e$, and inferred from the spacecraft potential, $N_{pot}$) increased from ~0.005 cm$^{-3}$ to ~0.01 cm$^{-3}$ at the density enhancement. At the trailing density ramp (between the vertical dotted lines), we observed a solitary electrostatic structure with a duration of ~1.4s (Figure 2c) without any accompanying magnetic field disturbances (Figure 2a). This structure exhibited an asymmetric bipolar $E_y$ component and a unipolar $E_x$ component in SSE coordinates (the $E_z$ component was unavailable due to unreliable calibration along the spacecraft spin axis). The bipolar $E_y$ established a potential hill, whereas the unipolar $E_x$, anti-parallel to the magnetic field (Figure 2a), resulted in a net electrostatic potential increase along the magnetic field line from the upstream region toward the density enhancement (i.e., the downstream region).

Due to the potential increase, electrons were heated as they entered the downstream region, resulting in broadened field-aligned velocity distribution functions (VDFs; $f(v_\parallel)$ ) with characteristic flat-top profiles (Figure 2d), as expected from simulations (An et al., 2025). Figure 3a compares time-averaged VDFs between the downstream and upstream regions (corresponding to the gray and red bars in Figure 2, respectively). After the upstream VDF is scaled upward by a factor of 2 to account for the density increase (red dotted line in Figure 3a), it closely overlaps with the downstream VDF (black line) in the anti-parallel direction. However, in the parallel direction considerable heating occurred, consistent with the anti-parallel orientation of the electrostatic field. Correspondingly, the electron energy in the parallel direction, $\int_0^{2.5e4} \frac{1}{2} m_e v_\parallel^2 f(v_\parallel) dv_\parallel / \int_0^{2.5e4} f(v_\parallel) dv_\parallel$, increased from ~50 eV in the upstream region to ~100 eV in the downstream region (Figure 2e).

This anti-parallel electric field may also decelerate incoming ions from the duskside in the parallel direction. Figures 3b-d show 2-D ion VDF slices in the BV plane (where the horizontal axis aligns with the magnetic field and the plane contains the ion bulk velocity) taken before, within, and after the density enhancement. Before and after the density enhancement (Figures 3b and 3d), the ion beams propagated at approximately -600 km/s and +360 km/s in the field-aligned direction (vertical dotted lines), respectively, while exhibiting E×B drift in the perpendicular direction. Within the density enhancement (Figure 3c), only the phase space density of the parallel ion beam was enhanced, indicating that the density increase resulted from deceleration of the parallel ion beam rather than overlapping of the two ion beams. This interpretation is also supported by Figure 1e, which shows that the energy flux enhancement corresponds to the parallel beam originating from the duskside.

Before and after the density enhancement, the ion distributions exhibit normal beam shapes with parallel anisotropy in Figure 3b and greater isotropy in Figure 3d. Thus, we expect that the ion distribution within the density enhancement in Figure 3c also exhibits a normal beam shape. However, due to the deceleration, part of the ion distribution approached the origin in velocity space, where the phase space volume of the ESA instrument is very small owing to its logarithmic energy channel configuration. In this volume, even one count can result in artificially high phase space densities (masked by the white circle in Figure 3c). Given the very low ion density within the lunar wake, this one-count level high statistical noise significantly corrupts the ion distribution near the origin, modulating the ion distribution shape around the white circle. This artifact makes it seem that only ions with larger perpendicular velocities (upper half of the distribution) were decelerated in the field-aligned direction to approximately 300 km/s (vertical dotted line). Although this is likely an artifact of statistics of low counts, we cannot fully rule out the possibility that it represents a real physical effect. If so, it would imply that field-aligned deceleration depends on perpendicular velocity, potentially due to varying electrostatic potential across magnetic field lines (see discussion in the supplementary material).

Overall, the electrostatic solitary structure observed within the trailing density ramp is consistent with an electrostatic shock, as predicted by simulations (An et al., 2025). It featured an electrostatic field pointing from downstream to upstream that heated electrons in the parallel direction and decelerated the incoming parallel ion beam causing density enhancement. We estimate three key parameters of this shock: the shock normal speed $V_s$ in the spacecraft frame, the spatial scale $d_s$, and the potential increase $\Delta\Phi$, as follows:

We assume the electrostatic shock was locally planar, with its normal direction along the magnetic field. For an ideal 1-D shock of infinite extent, P1 would traverse it at the shock normal speed $V_s$. However, if the shock plane has finite area, solar wind convection can cause P1 to encounter subsequently formed shocks across field lines before fully traversing the initial shock along field lines (see sketch in Figure S3 in the supplementary material). This effect reduces the effective spacecraft crossing speed of the shock. We model the shock crossing speed as a fraction of the shock normal speed $\alpha V_s$, where $\alpha$ is a free parameter ranging from 0 to 1. Accordingly, the spatial scale of the shock becomes $d_s \sim \alpha V_s \times 1.4 s$.

The potential change $\Delta\Phi$ can be estimated from: $-\int(E_x b_x + E_y b_y + E_z b_z)\alpha V_s dt$ over the 1.4s interval, where $E_i$ and $b_i$ are the three components of electric field and magnetic field unit vector, respectively, and the negative sign comes from field line direction pointing from upstream to downstream. Due to the absence of $E_z$ measurements, we estimate the contribution from the $E_z$ component to carry an uncertainty of ±50% and calculate $\Delta\Phi$ using only the $E_x$ and $E_y$ components. We thus obtain a potential increase across the shock $\Delta\Phi \sim 0.8 \pm 0.4\ mV \cdot s/m \times \alpha V_s$. The effects of this 50% uncertainty will be examined later. (Note that this potential estimate could include contributions from the background parallel electric field associated with the refilling process. Including its contribution is necessary when comparing with particle energy because particles experience the total electric field.)

Based on mass flux conservation, the upstream ion speed $V_{up}$ is approximately twice the downstream speed $V_{down}$ in the shock rest frame, and the corresponding decrease in ion kinetic energy across the shock is $\frac{3}{8} m_i V_{up}^2$. In the spacecraft frame, the upstream ion speed is determined from Figure 3d: $V_s + V_{up} \sim 360\ km/s$. Assuming the potential increase equals the ion kinetic energy decrease, i.e., $e\Delta\Phi = \frac{3}{8} m_i V_{up}^2$, we solve for $V_s$ as a function of $\alpha$.

For $\alpha = 1/4$, we obtain a shock normal speed of $V_s \sim 247\ km/s$, convecting with the duskside beam (the alternate solution of ~524 km/s is unphysical, outrunning the duskside beam), a potential increase of $\Delta\Phi \sim 50\ V$, and a spatial scale of $d_s \sim 86\ km$. This estimated potential increase is consistent with the observed electron parallel energy increase (Figure 2e). The ion speed decreased by ~57 km/s to ~303 km/s across the shock, which agrees well with the downstream ion distribution (upper half in Figure 3c). The local upstream Debye length $\lambda_e$ was ~1.54 km, and the electron skin depth $d_e$ was ~75 km, based on the electron temperature of 215 eV (Figure 1g) and density of 0.005 cm$^{-3}$ (Figure 2b). These values yield a shock thickness of $d_s \sim 56 \lambda_e$ or ~1.2 d$_e$, consistent with previous studies (e.g., Kato and Takabe, 2010; Dieckmann et al., 2014; An et al., 2025). Thus, the assumption of $\alpha = 1/4$ is reasonable. Larger values of $\alpha$ would result in smaller $V_s$ (down to ~200 km/s), larger $\Delta\Phi$ (up to ~160 V), and larger $d_s$ (up to ~280 km).

Next, we discuss the uncertainties associated with these estimates. If the density increase ratio exceeded 2 (e.g., due to measurement uncertainty or because only a portion of the ions were decelerated), the decrease in ion kinetic energy could approach $\frac{1}{2} m_i V_{up}^2$. In this extreme case, still assuming $\alpha = 1/4$, the shock speed $V_s$ increases slightly to $\sim 260 \ km/s$, yielding a shock potential increase of $\Delta\Phi \sim 52V$ and a spatial scale of $d_s \sim 59 \lambda_e$. Assuming that $E_z$ contributes a ±50% uncertainty, the shock normal speed $V_s$ ranges from ~227 km/s to 276 km/s, the potential $\Delta\Phi$ ranges from ~28 V to 68 V, and the spatial scale $d_s$ ranges from ~52 $\lambda_e$ to 63 $\lambda_e$. Additionally, the upstream beam speed may have been underestimated, as it was measured 1 minute away from the shock because of low ion counts immediately upstream of the shock. If the upstream speed was actually $V_s + V_{up} \sim 400 \ km/s$, then the shock normal speed $V_s$ becomes $\sim 310 \ km/s$, potential $\Delta\Phi$ becomes $\sim 62V$, and spatial scale $d_s$ becomes $\sim 70 \lambda_e$. Taking all sources of uncertainty into account, the estimated shock parameters remain within a physically reasonable range.

At the leading density ramp, no electrostatic shock was observed, and only electrostatic waves were present. It is possible that the leading shock had not yet formed at the time of observation. Alternatively, P1 may not have crossed the leading shock at all. If the shock normal speed in the spacecraft frame was sufficiently low, solar wind convection would dominate the spacecraft crossing. In this scenario, the observed leading edge of the downstream region could be a lateral boundary across field lines, rather than the interface with the anti-parallel ion beam along field lines. As shock structures developed and expanded over time, they would become more detectable by the spacecraft, as demonstrated in the next event.

**3.2. Event 2**

PIC simulations by An et al. (2025) show that after electrostatic shocks form, they start to dissipate and expand outward on both sides at speeds comparable to the ion acoustic speed. This process leads to the development of an extended downstream region replete with electrostatic structures, maintaining an enhanced electrostatic potential within it. We observe this later evolutionary stage in Event 2. Figure 4a shows a stable IMF dominated by the $B_y$ component, similar to Event 1. In the lunar wake, a large-scale density enhancement is evident between the vertical dotted lines in Figure 4c, corresponding to the thick red line in Figure 4i, which spans approximately 1000 km along the SSE-Y direction. Within this density enhancement, two ion beams were merging into a single population with reduced field-aligned speed (Figure 4e). Electrons were significantly heated, forming flat-top velocity distributions predominantly in the field-aligned direction (Figures 4f and 4g). Figure 4h reveals intense electrostatic wave activity with amplitudes of ~100 mV/m. Although we cannot identify a clear electrostatic solitary structure, asymmetric bipolar $E_y$ components transitioning from negative to positive (from parallel to anti-parallel relative to the IMF) across the density enhancement are evident. This pattern indicates net electrostatic fields pointing from downstream to upstream on both leading and trailing sides, creating a potential increase in the downstream region. These observations are consistent with characteristics of the later evolutionary stage in simulations by An et al. (2025).

Although P1 in this event was located at SSE-X ~ -2000 km, similar to its position in Event 1, it observed a more dissipated structure. This difference is attributed to the slower solar wind speed (~300 km/s) leading to a longer expansion time $t = \frac{L}{V_{sw}} \sim 6.7s$, compared to ~4s in Event 1

(where the solar wind speed was ~500 km/s). Given the upstream ion speed of ~200 km/s in the spacecraft frame (Figure 4e), the downstream ion speed of ~0 km/s, and a compression ratio of 2 (Figure 4c), we estimate the shock speed to be ~200 km/s. Thus, the shocks on two sides likely formed approximately 2.5 seconds earlier (calculated as ~1000 km / 2 / 200 km/s) at $t\sim4.2\ s$, which is comparable to the formation time in Event 1. Due to the rapid evolution (on the order of several seconds) and the small spatial scales of electrostatic shocks, capturing their early development as in Event 1 is rare. More commonly observed are their later evolutionary stages, as depicted in Event 2. Two additional examples are provided in the supplementary material.

## 4. Summary

In summary, using ARTEMIS observations, we identify an electrostatic shock at the center of the lunar wake in Event 1, where the two supersonic ion beams approaching from the two sides of the wake collided. This shock was characterized by an electrostatic solitary structure featuring a potential increase from upstream to downstream, which decelerated incoming ions leading to density enhancement (i.e., plasma compression) and heated electrons to form flat-top velocity distributions in the parallel direction. The cross-shock potential was estimated to be ~50 V, and the shock spatial scale was on the order of 50 Debye lengths. In Event 2, captured likely several seconds after the electrostatic shocks expanded from their initial formation, we observed a more dissipated structure with an extended downstream region (~1000 km), merging ion populations, and intense electrostatic waves. Net electrostatic fields oriented from downstream to upstream on both sides led to an enhanced potential and field-aligned electron heating in the downstream region. Our observations are in excellent agreement with predictions from PIC simulations by An et al. (2025). These processes may be applicable to plasma interactions with other airless celestial bodies.

**Acknowledgements**

TZL and XA are supported by the NASA grant 80NSSC23K0086. XA acknowledges NASA grant 80NSSC22K1634. We acknowledge THEMIS contract NAS5-02099. We acknowledge the SPEDAS team and NASA's Coordinated Data Analysis Web. We thank K. H. Glassmeier, U. Auster and W. Baumjohann for the use of the THEMIS/FGM data provided under the lead of the Technical University of Braunschweig and with financial support through the German Ministry for Economy and Technology and the German Center for Aviation and Space (DLR) under contract 50 OC 0302. We also thank the late C. W. Carlson and J. P. McFadden for use of THEMIS/ESA data.

**Open Research**

ARTEMIS dataset are available at NASA's Coordinated Data Analysis Web (CDAWeb, https://cdaweb.gsfc.nasa.gov/cgi-bin/eval1.cgi). The SPEDAS software (see Angelopoulos et al. (2019)) is available at https://themis.ssl.berkeley.edu/software.shtml.

# Figures

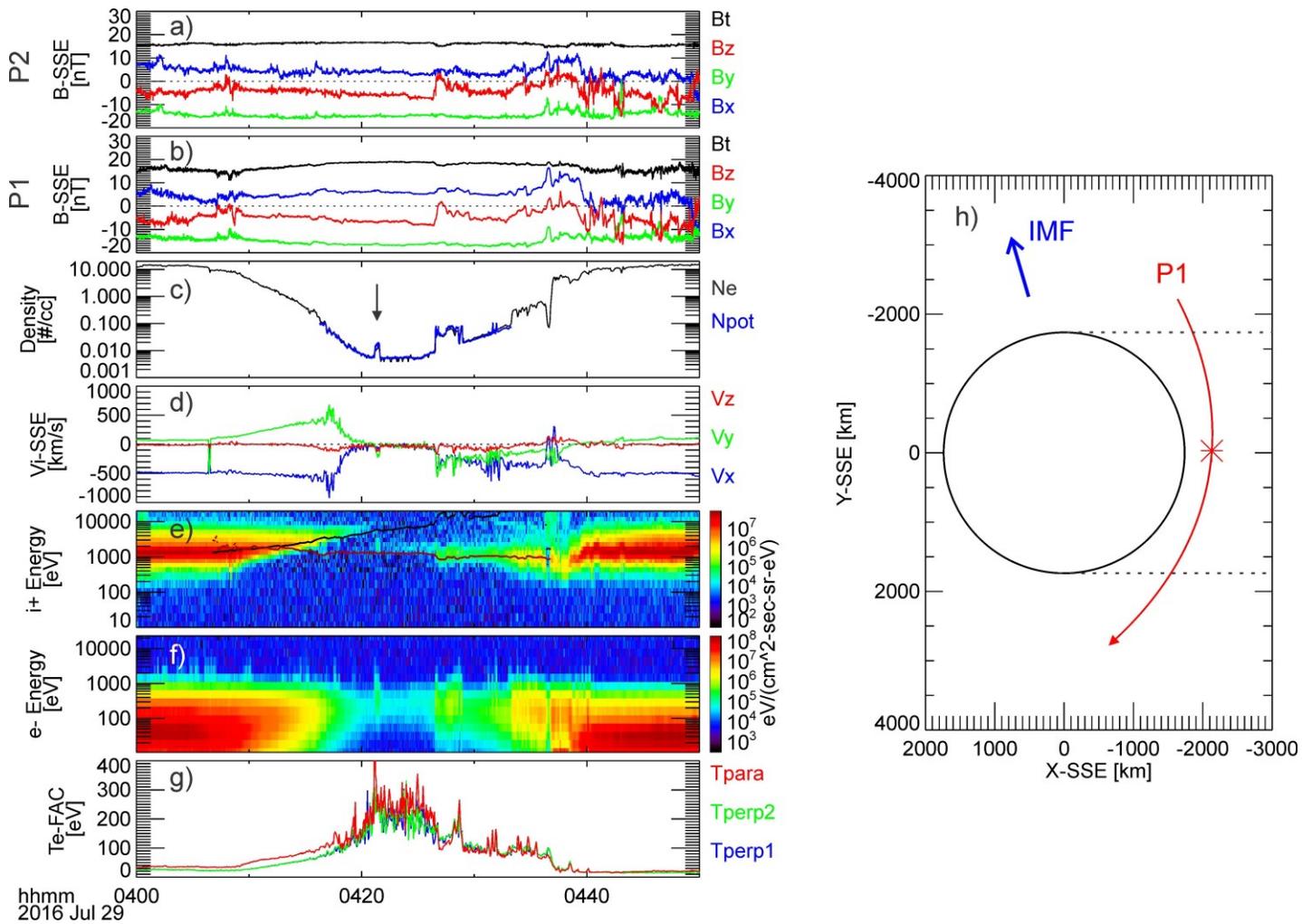

**Figure 1.** Overview of Event 1, showing (a) P2 observations of the background IMF, (b) P1 observations of magnetic field in SSE, (c) electron density measured by ESA (black) and inferred from spacecraft potential by fitting a Boltzmann relation (blue), (d) ion bulk velocity in SSE, (e) ion energy spectrum, with black and red solid lines indicating beam energies from the ion refilling model, (f) electron energy spectrum, and (g) electron temperature in field-aligned coordinates (FAC). Note that the very low ion speed within the wake is an instrumental artifact, because the very low ion density results in measurements near the one-count level which have large statistical noise. Panel (h) shows the trajectory of P1 (red) and the IMF orientation (blue) in the SSE-XY

plane. The star indicates the position where a density enhancement, discussed further in the text, was observed (black arrow in panel (c)). The two dotted lines indicate the lunar shadow.

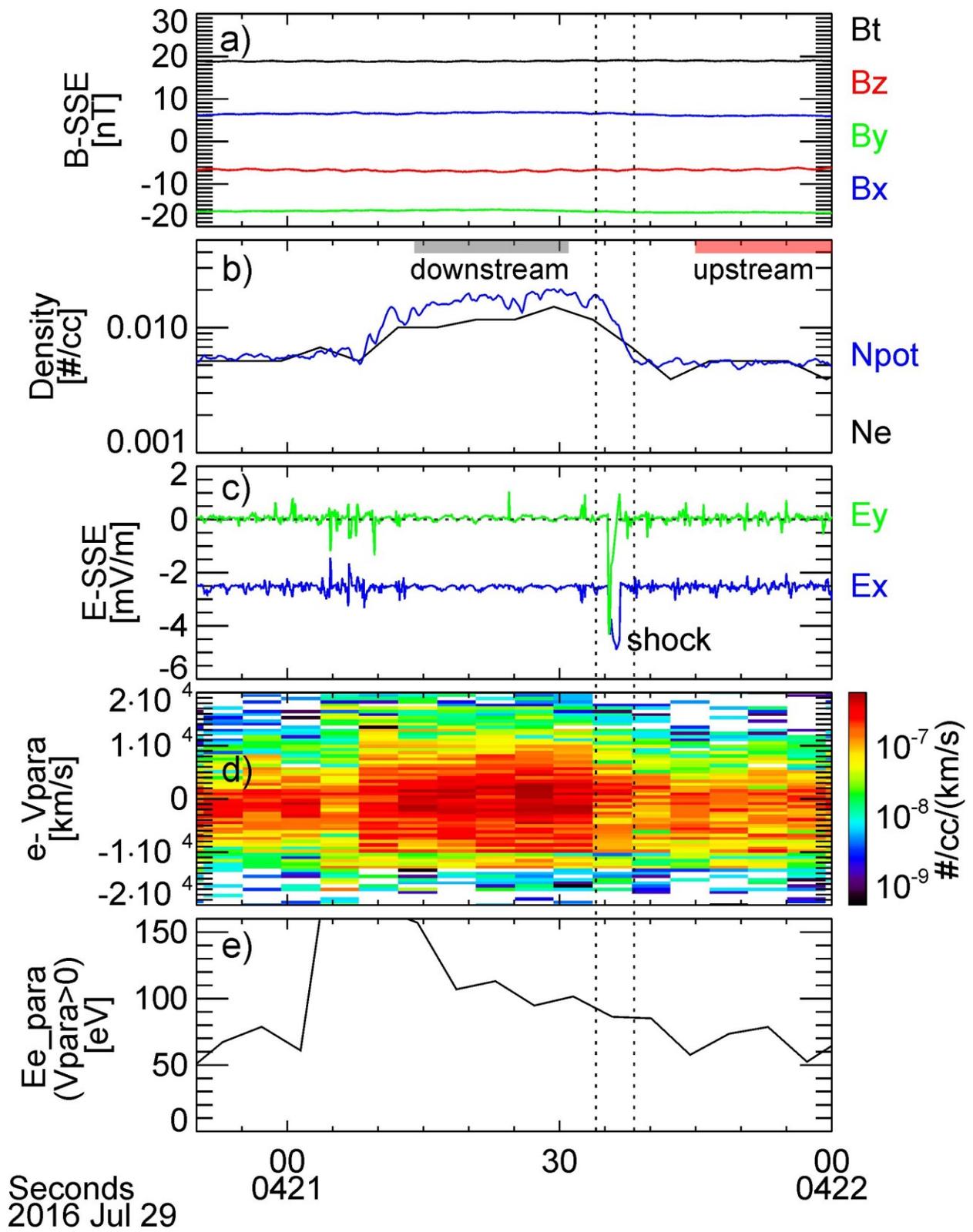

**Figure 2.** Zoom-in plot of the density enhancement. From top to bottom are (a) magnetic field in SSE, (b) electron density measured by ESA (black) and inferred from spacecraft potential (blue), (c) electric field in SSE (projected from the spacecraft spin plane, close to the equatorial plane; no Z component due to unreliable measurement along the spacecraft spin axis), (d) reduced 1-D electron velocity distribution function along the magnetic field by integrating over the perpendicular direction, (e) electron parallel energy from the parallel part of 1-D VDFs ($v_\parallel > 0$). Vertical dotted lines indicate the density ramp. Gray and red bars indicate the downstream and upstream time intervals used in Figure 3a.

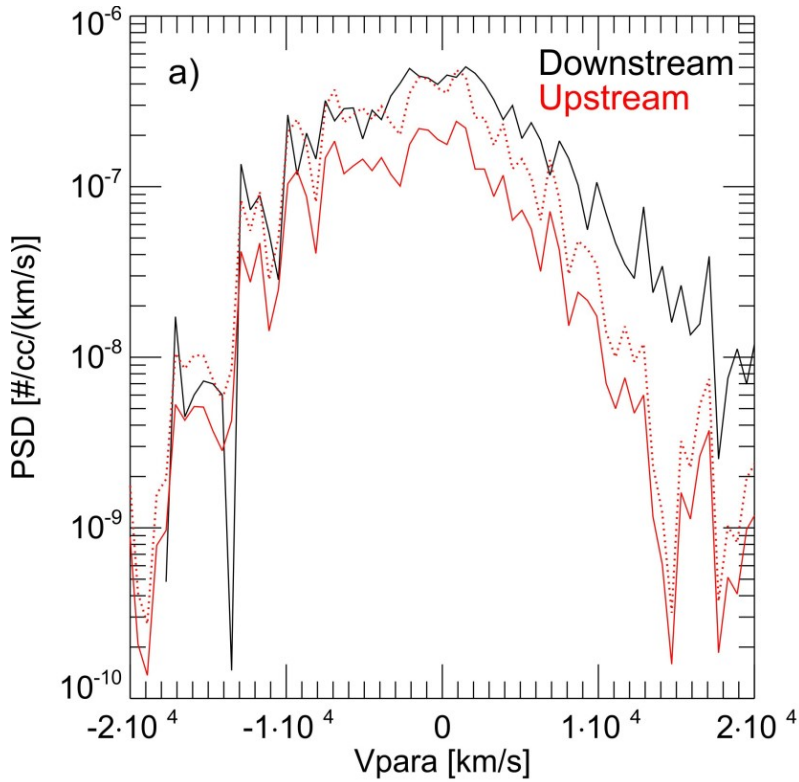

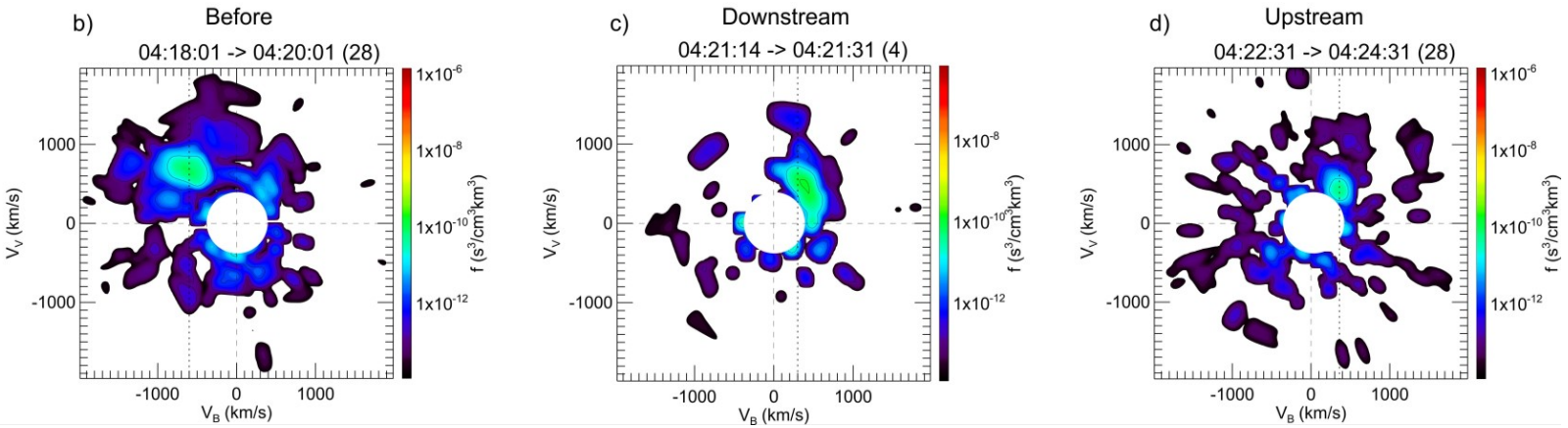

**Figure 3.** Electron and ion velocity distribution functions. Panel (a) shows reduced electron velocity distribution function along the magnetic field, averaged over 04:21:14-04:21:31 UT corresponding to the gray bar in Figure 2 (black line) and 04:21:45-04:22:45 UT corresponding to the red bar and beyond (red solid line). The red dotted line is the red solid line scaled upward by a factor of 2. Panels (b) to (d) are ion distribution slices in the BV plane where the horizontal axis aligns with the magnetic field and the plane contains the bulk velocity. Panels (b) and (d) are ion distribution functions 1 min before and after the density enhancement, because ion counts are too low immediately adjacent to it. Thus, the ion beam velocity in these panels slightly underestimates those adjacent to the density enhancement. The numbers in parentheses above each ion VDF indicate the number of slices averaged to reduce statistical noise. White circles mask the one-count noise region near the origin where phase space volume is very small. Vertical dotted lines in panels (b) to (d) indicate -600 km/s, 300 km/s, and 360 km/s, respectively.

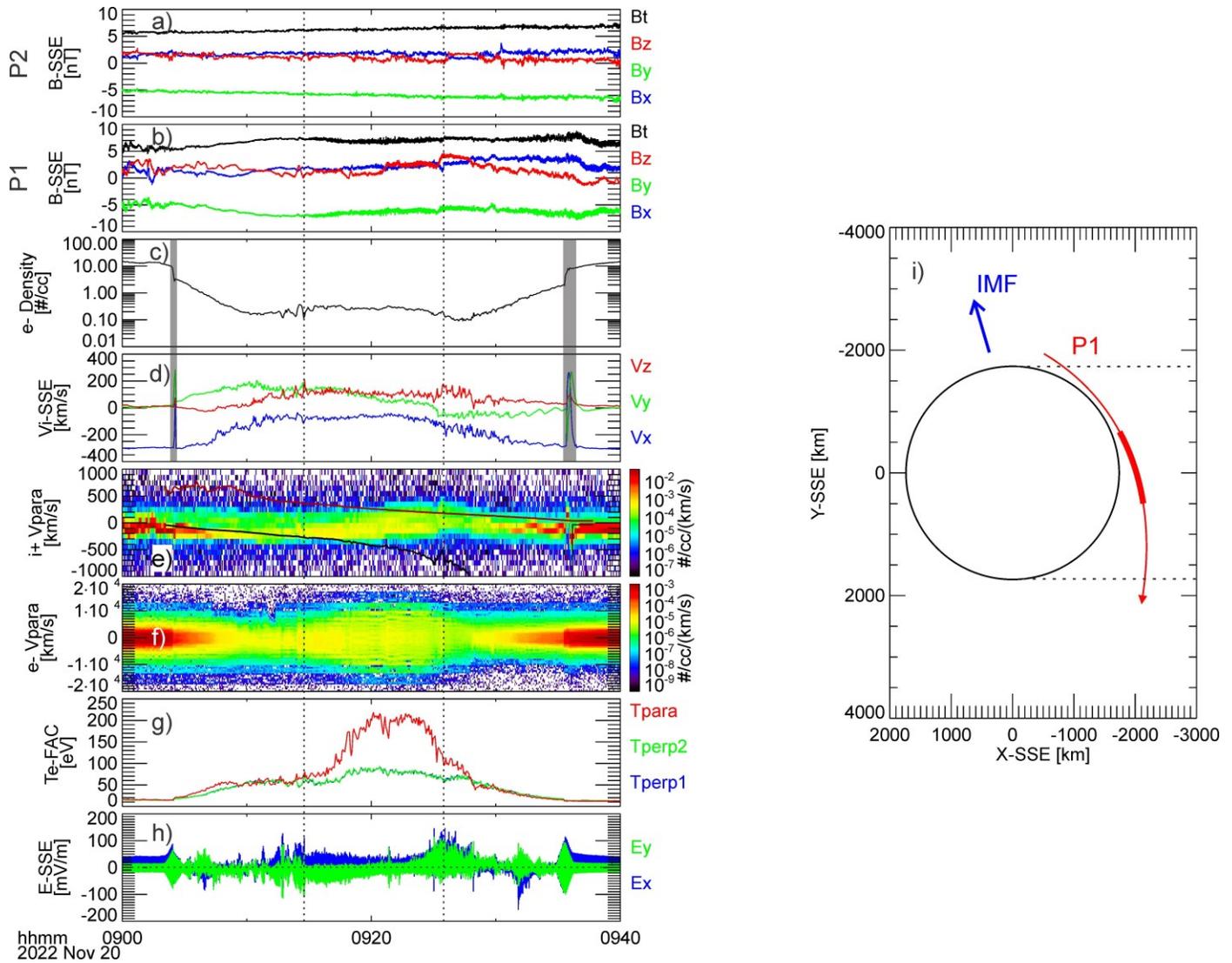

**Figure 4.** Overview of Event 2, showing (a) P2 observations of the IMF in SSE, (b) P1 observations of magnetic field in SSE, (c) electron density, (d) ion bulk velocity, (e) reduced ion VDF in the field-aligned direction. (f) reduced electron VDF in the field-aligned direction, (g) electron temperature in the field-aligned coordinates, and (h) electric field in SSE (projected from the spacecraft spin plane). Panel (i) is in a format similar to Figure 1h, and the thick line corresponds to the time interval between two vertical dotted lines. Density jumps and velocity spikes masked by shaded regions are instrumental artifacts due to the spacecraft potential affected by shadow entry/exit.

**Supplementary Material**

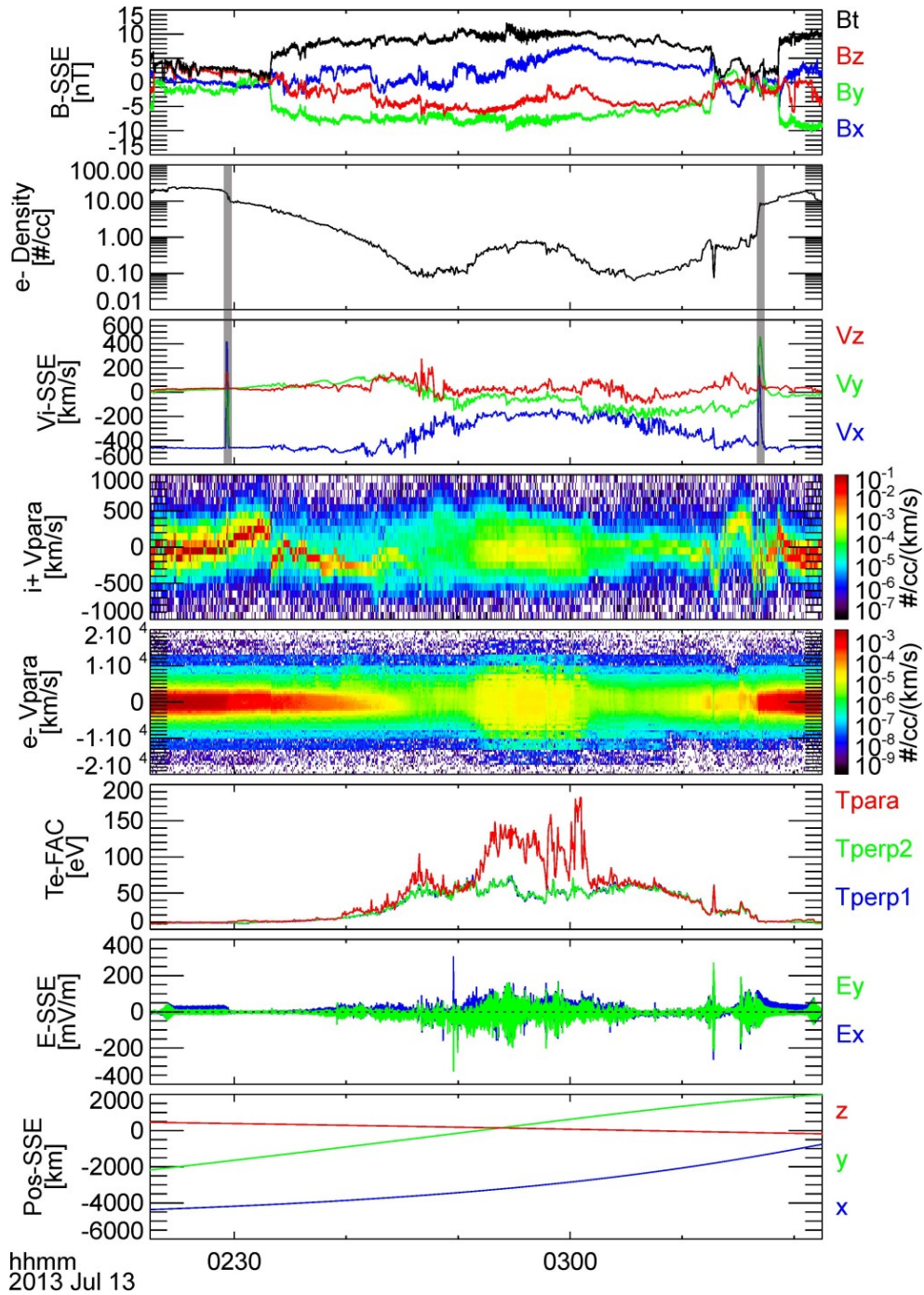

**Figure S1.** An observation example in a format similar to Figure 4, except last panel indicates spacecraft position. Similar to Event 2, a density enhancement was observed in the middle of the wake associated with electron heating in the parallel direction and bipolar $E_y$ pointing outward.

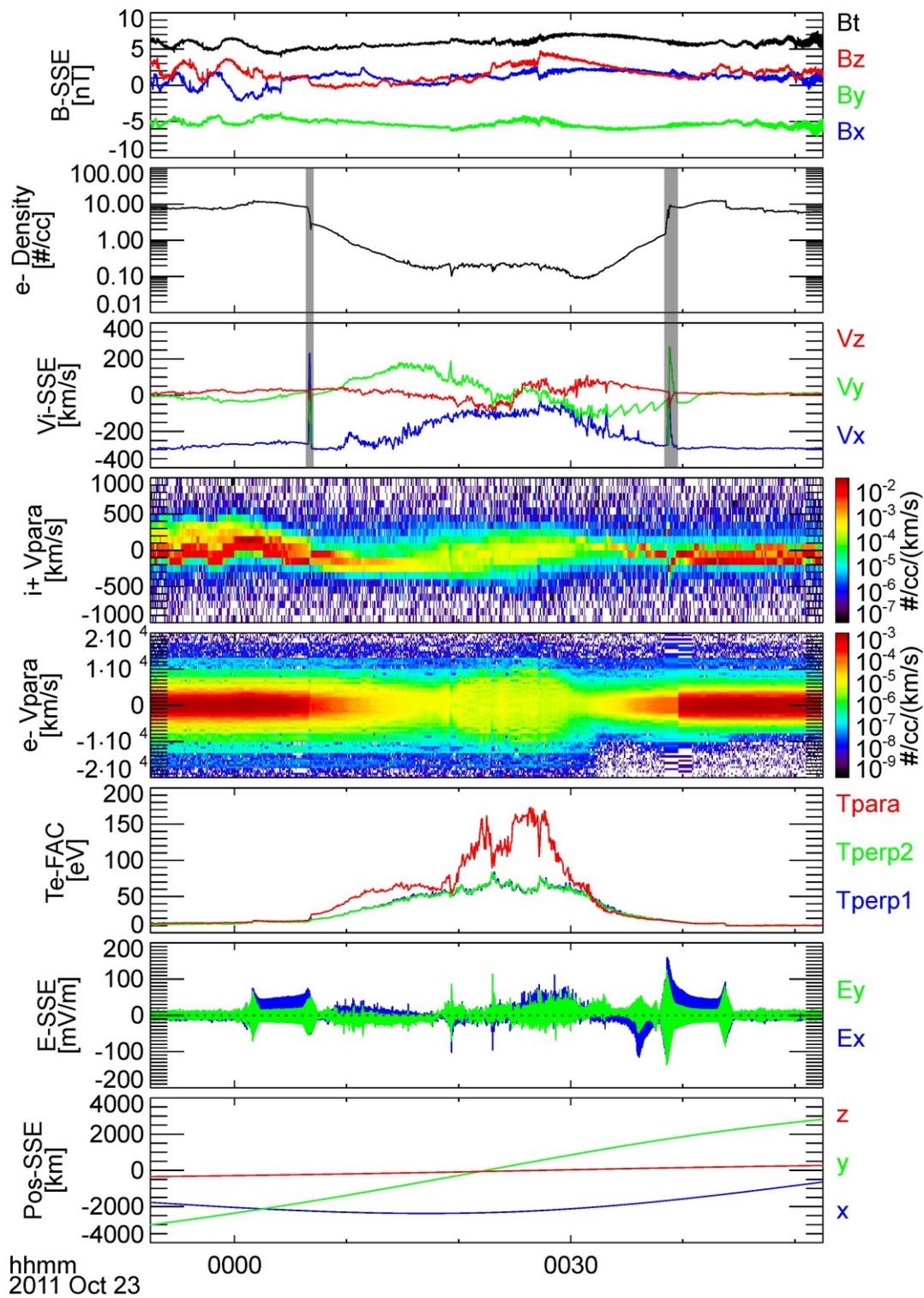

**Figure S2.** Another event in a format same as Figure S1.

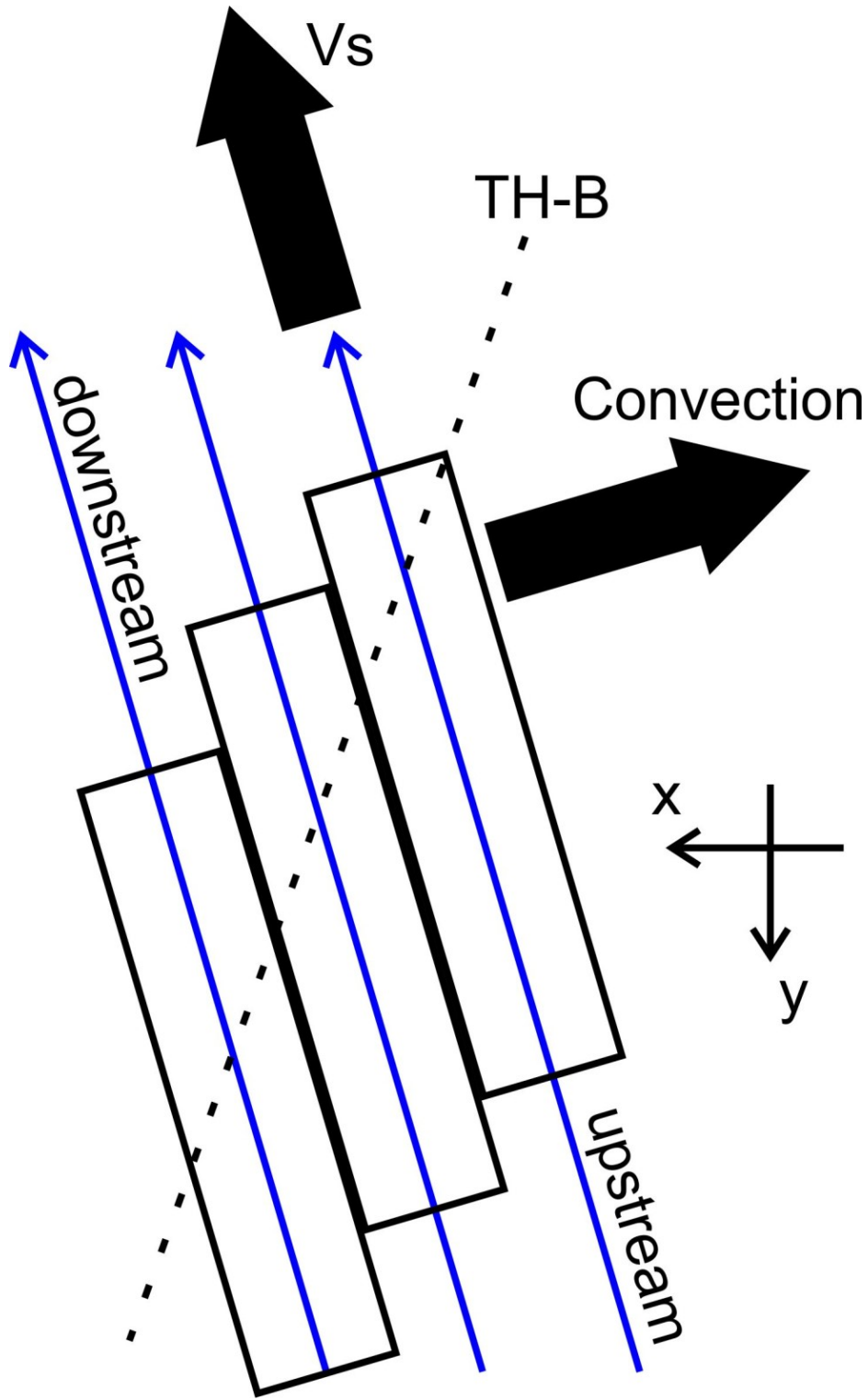

**Figure S3.** A sketch indicating shock crossing geometry. Black squares are shock structures at each field line (blue arrow), which are convecting perpendicular to field lines and propagating

parallel to field lines (black arrows). P1 thus crossed the shock structures along the dashed line and spent longer time than crossing a single shock structure with infinite area.

**Discussion about potential change across field lines**

In an ideal 1-D case, there is no potential change across field lines. However, spatial variations across field lines can occur due to the finite area of shock fronts (see Figure S3). Because the downstream regions had higher potential, these cross-field variations can generate an electrostatic field component perpendicular to the magnetic field in the anti-sunward direction, possibly corresponding to the perpendicular component of the unipolar $E_x$ in Figure 2c. Additionally, due to the cylindrical shape of the lunar wake boundaries, spatial variations may also occur in the SSE-Z direction.

For gyrospeed of 100 km/s, the ion gyroradii are ~50 km, comparable to the shock spatial scale. Thus, the guiding centers of ions from the upper and lower halves of the ion distribution in Figure 3c can be separated by several hundred kilometers, mainly in the SSE-Z direction. This separation may be sufficient to produce noticeable energy differences.

Figure 3c shows an asymmetric bipolar $E_y$. This asymmetry may also arise from spatial variation across field lines. The leading negative $E_y$ peak could be associated with a shock that formed earlier than the trailing positive $E_y$ peak, resulting in a stronger amplitude. These 3-D effects increase the complexity of comparisons with 1-D models.

# References

An, X., Angelopoulos, V., Liu, T. Z., Artemyev, A., Poppe, A. R., & Ma, D. (2025). Plasma Refilling of the Lunar Wake: Plasma-Vacuum Interactions, Electrostatic Shocks, and Electromagnetic Instabilities. Journal of Geophysical Research: Space Physics, 130(7). https://doi.org/10.1029/2025ja034205

Angelopoulos, V. (2008), The THEMIS mission, Space Sci. Rev., 141, 5–34, doi:10.1007/s11214-008-9336-1.

Angelopoulos, V., Cruce, P., Drozdov, A. et al. The Space Physics Environment Data Analysis System (SPEDAS) [software]. Space Sci Rev (2019) 215: 9. https://doi.org/10.1007/s11214-018-0576-4

Colburn, D. S., R. G. Currie, J. D. Mihalov, and C. P. Sonett (1967), Diamagnetic solar-wind cavity discovered behind moon, *Science*, **158**, 1040–1042.

Davidson, R. C., Krall, N. A., Papadopoulos, K., & Shanny, R. (1970). Electron Heating by Electron-Ion Beam Instabilities. Physical Review Letters, 24(11), 579–582. https://doi.org/10.1103/physrevlett.24.579

Denavit, J. (1979), Collisionless plasma expansion into a vacuum. Phys. Fluids **22**, 1384–1392

Dieckmann, M. E., Sarri, G., Doria, D., Ahmed, H., & Borghesi, M. (2014). Evolution of slow electrostatic shock into a plasma shock mediated by electrostatic turbulence. New Journal of Physics, 16(7), 073001. https://doi.org/10.1088/1367-2630/16/7/073001

Farrell, W. M., M. L. Kaiser, and J. T. Steinberg (1997), Electrostatic instability in the central lunar wake: A process for replenishing the plasma void?, *Geophys. Res. Lett.*, *24*, 1135–1138

Forslund, D. W., & Freidberg, J. P. (1971). Theory of Laminar Collisionless Shocks. Physical Review Letters, 27(18), 1189–1192. https://doi.org/10.1103/physrevlett.27.1189

Forslund, D. W., & Shonk, C. R. (1970). Formation and Structure of Electrostatic Collisionless Shocks. Physical Review Letters, 25(25), 1699–1702. https://doi.org/10.1103/physrevlett.25.1699

Halekas, J. S., S. D. Bale, D. L. Mitchell, and R. P. Lin (2005), Magnetic fields and electrons in the lunar plasma wake, *J. Geophys. Res.*, *110*, A07222, doi:10.1029/2004JA010991

Halekas, J. S., A. R. Poppe, and J. P. McFadden (2014), The effects of solar wind velocity distributions on the refilling of the lunar wake: ARTEMIS observations and comparisons to one-dimensional theory, *J. Geophys. Res. Space Physics*, 119, 5133–5149, doi:10.1002/2014JA020083.

Karimabadi, H., Omidi, N., & Quest, K. B. (1991). Two-dimensional simulations of the ion/ion acoustic instability and electrostatic shocks. Geophysical Research Letters, 18(10), 1813–1816. https://doi.org/10.1029/91gl02241

Kato, T. N., & Takabe, H. (2010). Electrostatic and electromagnetic instabilities associated with electrostatic shocks: Two-dimensional particle-in-cell simulation. Physics of Plasmas, 17(3), 032114. https://doi.org/10.1063/1.3372138

Mozer, F. S. (1981). ISEE-1 observations of electrostatic shocks on auroral zone field lines between 2.5 and 7 Earth radii. Geophysical Research Letters, 8(7), 823–826. https://doi.org/10.1029/gl008i007p00823

Sakawa, Y., T. Morita, Y. Kuramitsu, H. Takabe, Collisionless electrostatic shock generation using high-energy laser systems, Advances in Physics: X, 10.1080/23746149.2016.1213660, **1**, 3, (425-443), (2016).

Samir, U., K. H. Wright Jr., and N. H. Stone (1983), The expansion of a plasma into a vacuum: Basic phenomena and processes and applications to space plasma physics, *Rev. Geophys.*, 21(7), 1631–1646, doi:10.1029/RG021i007p01631.

Stringer, T. E. (1964). Electrostatic instabilities in current-carrying and counterstreaming plasmas. Journal of Nuclear Energy. Part C, Plasma Physics, Accelerators, Thermonuclear Research, 6(3), 267–279. https://doi.org/10.1088/0368-3281/6/3/305

Zhang, H., et al. (2012), Outward expansion of the lunar wake: ARTEMIS observations, *Geophys. Res. Lett.*, *39*, L18104, doi:10.1029/2012GL052839.

Zhang, H., K. K. Khurana, M. G. Kivelson, V. Angelopoulos, W. X. Wan, L. B. Liu, Q.-G. Zong, Z. Y. Pu, Q. Q. Shi, and W. L. Liu (2014), Three-dimensional lunar wake reconstructed from ARTEMIS data, *J. Geophys. Res., 119*, 5220–5243, doi:10.1002/2014JA020111.